\documentclass[aps,physrev,preprint,groupedaddress]{revtex4-2}

\usepackage{graphicx}
\usepackage{tikz}
\usepackage[hidelinks]{hyperref}
\usepackage{orcidlink}

\begin{document}

\preprint{PRE-PRINT}

\title{\textbf{Experimental investigation of ridge-induced secondary motions in turbulent channel flows} 
}%

\author{Mattias Nilsson-Takeuchi \orcidlink{0000-0002-2071-1075}}
\email[]{Contact author: M.Nilsson-Takeuchi@soton.ac.uk}

\author{Bharathram Ganapathisubramani \orcidlink{0000-0001-9817-0486}}%
\affiliation{%
 Department of Aeronautical and Astronautical Engineering, University of Southampton, Southampton, SO17 1BJ, UK\\}%



\date{\today}

\begin{abstract}

Many engineering and environmental surfaces exhibit spatial heterogeneity in the spanwise direction and encompass multiple surface length scales. When the dominant spanwise length scale is on the order of the largest flow scales (e.g., the boundary layer thickness or channel half-height, $\delta$), localized $\delta$-scale secondary flows can form \citep{Vanderwel2015}. These secondary-flow generating heterogeneous surfaces are typically classfied as ``ridge-type'' (with spanwise variation in surface elevation) and ``strip-type'' (spanwise variation in skin-friction including through variation in roughness characteristics). Both types of surfaces have been explored in previous studies at high Reynolds numbers using experiments with focus on a variety of characteristics such as mean flow, turbulent statistics and structure while lower Reynolds number studies in channels have examined influence on skin-friction in addition to flow characteristics/mechanisms.  In this work, we augment the previous work through experiments on ridge-type roughness in turbulent channel flows over a large range of Reynolds numbers.  Our findings reveal that the global skin friction, as inferred through pressure drop in the channel along the streamwise direction, of these surfaces exhibits a log-linear asymptotic behaviour that is characteristic of a fully-rough surface at sufficiently high Reynolds numbers. Yet, the roughness function where this is observed is ``low'' compared to other established homogeneous rough surfaces. Moreover, the mean flow, turbulent statistics, and two-point correlation analyses indicate that the secondary flow structure, spatial extent, and magnitude remain largely invariant over the Reynolds number range considered. These findings have important implications on predicting drag of these heterogeneous surfaces as well as developing new models for the flow structure. 
\end{abstract}


\maketitle

\section{Introduction}\label{sec:intro}

\begin{figure}[!htb]
  \centering
  \includegraphics[width=\textwidth]{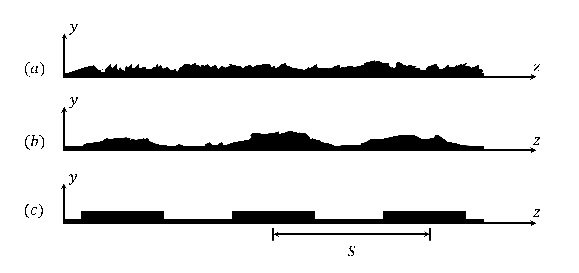}
  \caption[Illustration of spanwise heterogeneity.]{Illustration of spanwise heterogeneity. (a) A homogeneous rough surface; (b) a spanwise heterogeneous rough surface; (c) a simplified canonical version typically implemented in research. Here, $S$ denotes the spanwise wavelength.}
  \label{Figure:HetSurf}
\end{figure}

Spanwise heterogeneous rough surfaces have received increasing attention over the past few decades due to their common appearance in many of the practical engineering applications and environmental flows. Examples include turbine blades damaged by sand ingestion \citep{Barros2014, Barros2019}, herringbone riblets \citep{Nugroho2013} and bio-fouled ship hulls \citep{Schultz2007bio}. In environmental flows, flows over riverbeds with non-uniform sediment distribution and atmospheric flows over urban areas, plant canopies, or wind farms \citep[see for example][]{Nezu1993} also display such heterogeneity. Unlike homogeneous rough surfaces, spanwise heterogeneous surfaces are characterized by a spatially periodic variation with a distinct spanwise wavelength $S$, see Figure \ref{Figure:HetSurf}. 

A primary consequence of such spanwise heterogeneity is the generation of low-energy secondary flows in the cross-stream plane. Although their strength is only a few percent of the primary flow, these large-scale motions can substantially alter the mean velocity profile, turbulent statistics, and thereby influence drag and heat transfer throughout the boundary layer \citep[see for example][]{Medjnoun2018, Vanderwel2019}. These secondary flows are inherently three-dimensional, exhibiting alternating high-momentum (HMP) and low-momentum (LMP) regions, with corresponding downwash and upwash events that lead to a heterogeneous distribution of turbulent stresses. Because these motions are driven by turbulence imbalances, they are classified as Prandtl's secondary flows of the second kind \citep{Hinze1973, Anderson2015, Hwang2018}.

For clarity and control in parametric studies, heterogeneous surfaces are often idealized into two canonical types: \textit{ridge-type} and \textit{strip-type} surfaces. Ridge-type surfaces exhibit a spanwise variation in surface elevation. Typically, the elevated regions (ridges) and the lower, flat regions (valleys) are smooth, leading to a form of “heterogeneous roughness” that does not induce additional form drag in the conventional sense \citep{Castro2021}. In contrast, strip-type surfaces feature alternating regions of high and low surface skin friction, often achieved by alternating smooth and rough patches \citep{Hinze1967, Turk2014}. Recent studies have shown that the spanwise wavelength $S$ is a key determinant of the secondary flow behaviour \citep{Vanderwel2015}. Moreover, additional parameters --- such as the cross-sectional shape of the ridges, their width, valley (or gap) width, and ridge height $h$ --- play important roles in defining the structure and intensity of the secondary flows \citep{Medjnoun2020, Zampiron2020, Yang2018}. For instance, while early studies assumed ridge width to be of minor importance, subsequent work has demonstrated that both shape and width significantly affect secondary and even tertiary flow structures. State-of-the-art experimental campaigns \citep[e.g.,][]{Medjnoun2020, Zampiron2020, Wangsawijaya2020} and recent DNS studies \citep[e.g.,][]{Castro2021, Castro2024, Zhdanov2024} have provided valuable insights into the dynamics of ridge-induced secondary flows. However, these studies have been limited by either the range of Reynolds numbers or the flow configurations investigated. Additionally, emerging rapid predictive tools such as linearised RANS models \citep{Zampino2022} offer promising avenues for exploring geometries that are impractical to manufacture, further motivating targeted experimental studies.

Experimental studies by \citet{vonDeyn2021} and \citet{Frohnapfel2024} have shown that the skin friction behaviour over spanwise-heterogeneous surfaces deviates from that of smooth-wall canonical turbulent boundary layers. Specifically, their results suggest that such flows do not exhibit statistical equilibrium or Reynolds number–independent skin friction. However, as these studies were conducted at moderate Reynolds numbers, it remains an open question whether classical scaling behaviour might emerge at higher Reynolds numbers. Increased attention has also been given to the unsteady dynamics of secondary flows. Notably, the work by \citet{Zampiron2020}, \citet{Wangsawijaya2020} and \citet{Wangsawijaya2022} have shown that these flows are not strictly steady or spatially fixed. Instead, they exhibit spanwise meandering, energy redistribution across scales, and modulations in turbulence structure. These findings, drawn from both ridge-type and strip-type configurations, underscore the complex and time-dependent nature of secondary flows induced by spanwise heterogeneity.

Despite recent advances, many aspects of secondary flows remain unresolved. In particular, most experimental studies have focused on boundary layers or open-channel flows or have been limited to a low Reynolds numbers. In addition, the influence of key geometric parameters, such as ridge shape, width, valley spacing, and height, on the formation and strength of secondary flows is not yet fully understood. In this work, we address some of these gaps by performing a systematic experimental investigation of ridge-induced secondary flows in a closed-channel flow over a range of Reynolds numbers ($Re_\tau = \delta U_\tau/\nu$ up to 4000). In doing so, we aim to quantify the impact of surface heterogeneity on skin-friction and its scaling with Reynolds number. We also aim to examine how the mean flow and turbulent statistics are modulated by secondary flows. Finally, we aim to characterize the spatial coherence and orientation of turbulent structures using two-point correlation analyses. By extending the investigation to a wider range of Reynolds numbers in a closed-channel flow, this work not only provides new experimental data but also deepens our understanding of how ridge-induced secondary flows.

The paper is structured as follows: Section \ref{sec:methods} details the experimental methodology, including the water channel facility, ridge geometry, skin friction measurements, and the stereoscopic Particle Image Velocimetry (PIV) setup. Section \ref{sec:results} presents the experimental results, covering skin friction behaviour, mean flow modulation, turbulent statistics, and turbulent structure analysis via two-point correlations. Finally, Section \ref{sec:conclusions} summarizes the key findings.

\section{Experimental setup and method}\label{sec:methods}

\subsection{Facility and geometry}\label{sec:waterchannel}

\begin{figure}
    \centering
    \includegraphics[width=\textwidth]{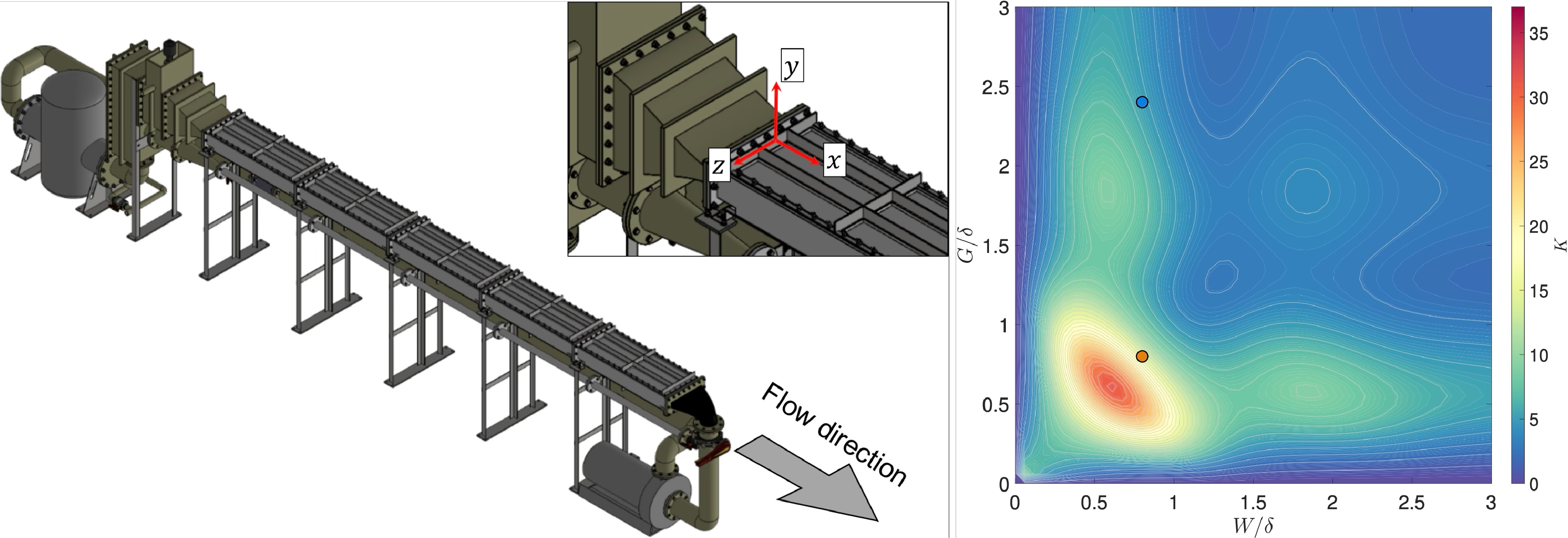}
    \caption{(left) Schematic of channel flow and (right) Volume averaged kinetic energy contours recreated from \citet{Zampino2022}. Selected ridge width $W$ and spacing $S = W + G$ marked on top for $S/W = 2$ (orange marker) and $S/W = 4$ (blue marker).}
    \label{fig:exp}
\end{figure}

A closed water channel facility with a 6125 mm long test section and a wetted cross section of width ($W$) $400\ mm$ and full height ($H$) $50\ mm$ was used to carry out the experiments, see Figure \ref{fig:exp}(left). The relevant fluid dynamic length scale for this flow is the channel half-height and this is represented as $\delta$. The cross-sectional dimensions yield an aspect ratio of 8, allowing the centerline flow to be considered two-dimensional when the flow is fully developed \citep{Monty2005}. The coordinate system is chosen such that $x$, $y$ and $z$ are in the streamwise, wall-normal, and spanwise direction, respectively. The test section duct is divided into five sections, each 1225 mm long. The bottom and top surfaces of each channel duct section are removable and can be swapped out for any surface of interest. The uncertainty of the spanwise alignment of the plate is $\pm 1\ mm$.  

The flow is driven by a 30 kW pump with a feedback controller that allows setting the bulk velocity $U_b$ or the bulk Reynolds number $Re_b = U_b H/\nu = 2U_b\delta/\nu$  to within $0.5\%$ of the target value. The flow rate is measured using a Yokogawa AXF150 electromagnetic flow meter with a standard accuracy of 0.35\% of the reading value. The water temperature is measured using a standard IEC-751 class B PT100 resistance platinum probe in a 4-wire connection, having a measurement error of approximately $\pm(0.3 + 0.005*T)$ degrees Celsius, where $T$ is the measured temperature.  A microcontroller with a 10-bit A/D converter was used to sample the flow meter and temperature probe. A typical temperature during a run was between 16 and 20 degrees Celsius. The temperature measurements were used to determine the water density $\rho$ and viscosity $\mu$ through empirical data fits to the data \cite{Tanaka2001,Cengel2010}.



With the long list of secondary flow studies with various surface configurations presented in Section \ref{sec:intro} there is plenty of inspiration to draw from when choosing the surface geometry for generating the secondary flows. There are a range of shapes, ridge widths, ridge heights and ridge spacings to consider. Rectangular ridges, in addition to their simplicity and ease of manufacturing, have been shown to produce strong localised secondary flows (see for example \citet{Medjnoun2018, Medjnoun2020}). As such, rectangular ridges provides good conditions for identification and characterisation of these flows and were selected for this work. \citet{Zampino2022} recently used a linearised Spalart-Allmaras model to simulate ridge-induced secondary flows and observed high volume averaged kinetic energy in two regions of combinations of ridge width $W$ and valley width $G$. From these results, two surface configurations were chosen in regions where the secondary flows are expected to be the most energetic, see \ref{fig:exp}(right). The spanwise ridge width $W$ was kept constant at $0.8\delta$ while two different spanwise spacings (and therefore valley widths), $S = 1.6\delta$ and $S = 3.2\delta$, were used. From here on these cases will be referred to as $S/W = 2$ and $S/W = 4$, respectively. The ridge height was chosen to be $k = 2.5mm = 0.1\delta$ to provide a sufficient spanwise height-step to induce the secondary flows \cite{Willingham2014}.  The channel top and bottom test surfaces in all five test sections were covered with the ridges aligned with the streamwise direction (over $120H$ of development length that is sufficient for fully-developed channel flow).


\subsection{Skin friction measurements}\label{sec:pressuretransducer}

Because the important scaling parameter $U_\tau$ is determined by the wall skin friction $\tau_w$, skin friction is a crucial quantity when characterising turbulent flows over both smooth and rough surfaces. In addition, surface drag is a performance metric of much interest which is also determined from the wall skin friction. For the canonical fully developed channel flow, it can be shown from the boundary layer equations that

\begin{equation}\label{eq:skin-friction}
    \tau_w = -\frac{H}{2}\frac{dp}{dx} = -\delta \frac{dp}{dx}
\end{equation}

The wall shear stress in fully developed channel flow can therefore be determined through equation \ref{eq:skin-friction} if accurate measurements of the streamwise pressure gradient is available. The skin friction is then commonly normalised by the dynamic pressure and presented as a skin friction coefficient, defined as

\begin{equation}\label{eq:skin-friction-norm}
    C_f = \frac{\tau_w}{\frac{1}{2}\rho U_b^2} = 2\left(\frac{U_\tau}{U_b} \right)^2
\end{equation}

where $U_b = Q/A$ is the bulk velocity, calculated from the flow rate $Q$ and test section cross-sectional area $A$. It should be noted here that the channel half-height becomes non-trivial in equation \ref{eq:skin-friction} when roughness is introduced. If the roughness height is significantly smaller than the half-height (i.e. $h << \delta$), then it is trivial to use $\delta$ as the half-height. However, when $h <\delta$ (in our case and most other studies in the literature $h \sim 0.1\delta$), then we need to clearly define the length-scale used to determine the skin-friction in equation \ref{eq:skin-friction}. The half-height $H/2$ can then be interpreted as the distance from the centerline to where the effective wall-shear stress applies as if acting on a virtual flat plate above the base smooth surface. However, the distance to this virtual flat surface is arbitrary and can be chosen in several ways. For example, due to the aim of their study, \citet{vonDeyn2021} chose the channel height of ridge-type surfaces such that the surface structures did not induce any drag variation under laminar flow conditions. Another common choice due to its easy accessibility is the average ``meltdown" height where the average height of the surface protrusions is subtracted from the channel half-height to represent an average location of a virtual flat wall if the surface elements were ``melted" down. In this study the average meltdown half-channel height was used for all ridge-type surfaces in quation \ref{eq:skin-friction}. 

To measure the streamwise pressure gradient in the channel, a 5 psi ($\approx 35kPa$) SCANIVALVE DSA3307-PTP differential pressure scanner with a 16 bit A/D converter, providing pressure measurements at up to 850 Hz on 8 channels (dependent on the total data output sent over the $100\ Mbit/s$ Ethernet connection), was used. The accuracy of the scanner is 0.5\% of the full scale measurement range (0.5\% of 5 psi $\approx 172\ Pa$). The scanner was then connected to a series of 8 static pressure tappings installed along the centerline of the channel side wall, similarly to \citet{Schultz2013, vonDeyn2021}. The static pressure tappings were $3\ mm$ in diameter, located approximately $8H$ apart (evenly spaced in such a way that the middle tap in each test section is centred in the streamwise direction) to give a sufficiently large pressure difference that the pressure scanner can accurately measure it. All pressure measurements were then corrected due to for example finite diameter effects using the static pressure tap correction introduced by \citet{McKeon2002} and has a very minimal effect on the results. Pressure measurements were acquired over 28000 ($\delta/U_b$) convection cycles through a bulk Reynolds number sweep from $Re_b = 35000$ to $Re_b = 190000$ in steps of $5000$. The pressure gradient was determined by fitting a straight line through the data points of the static pressure. Each measurement was repeated six times and averaged over the repeats to obtain the final skin friction coefficient.

\subsection{Stereoscopic Particle Image Velocimetry}\label{sec:channel_PIV}

Cross-stream stereo PIV was used to capture the velocity field in a spanwise---wall-normal plane with the laser sheet located at $x_m \approx 5.5 m = 110H$ downstream of the channel inlet. A {\it Litron Bernoulli} Nd:YAG Laser ($200 mJ$) was used to illuminate polyamide particles $55\mu m$ in diameter. The laser beam is passed over two mirrors and through sheet-expander optics to produce an approximately $1\ mm$ thick laser sheet entering from the sidewall of the channel, covering the whole channel section height. Two {\it LaVision ImagerProLX} 16 MP cameras with 200 mm {\it Nikon} lenses angled $\pm45^{\circ}$ degrees to the laser sheet, attached via Scheimpflug adapters, were used to capture the particle displacements in a field of view (FOV) approximately $90\times60 mm^2$  ($z\times y$), allowing at least one whole spanwise wavelength and the entire channel height to be captured for all surfaces considered (see details on the ridge geometry below). $45^{\circ}$ water prisms were constructed to avoid image distortions due to the refraction index change between the ambient air and the water in the channel. Self-calibration was used to account for misalignment of the calibration target with the laser sheet.  Sets of 3000 image pairs were captured at 0.6 Hz for each test case and processed using a multi-pass scheme down to $24\times 24\ px^2$ interrogation windows with 50\% overlap, resulting in a vector spacing of approximately $0.32\times 0.32\ mm^2$. For both surface configurations images were captured at three target friction Reynolds numbers, $Re_\tau \approx 2000,\ 3000$ and $4000$. 


%

Because the field of view  of the PIV experiments were kept constant, the vector spacing is the same for all cases in physical space. However, in inner-scaled units the $y^+$, $z^+$ coordinate system is a function of skin friction velocity and viscosity which both varies with Reynolds number. Hence, the inner-scaled resolution effectively varies across cases (worsening for higher Reynolds numbers). Therefore, any inner-scaled velocity fields or profiles presented here have been filtered to the worst resolution case ($\approx 120\ \frac{u_\tau}{\nu}$, i.e. 120 wall units) using a Gaussian kernel convolution to account for any differences in turbulent stresses resulting from the variation in grid resolution. 


\section{Results}\label{sec:results}

\subsection{Surface skin friction}

\begin{figure}
    \centering
    \includegraphics[width=0.8\textwidth]{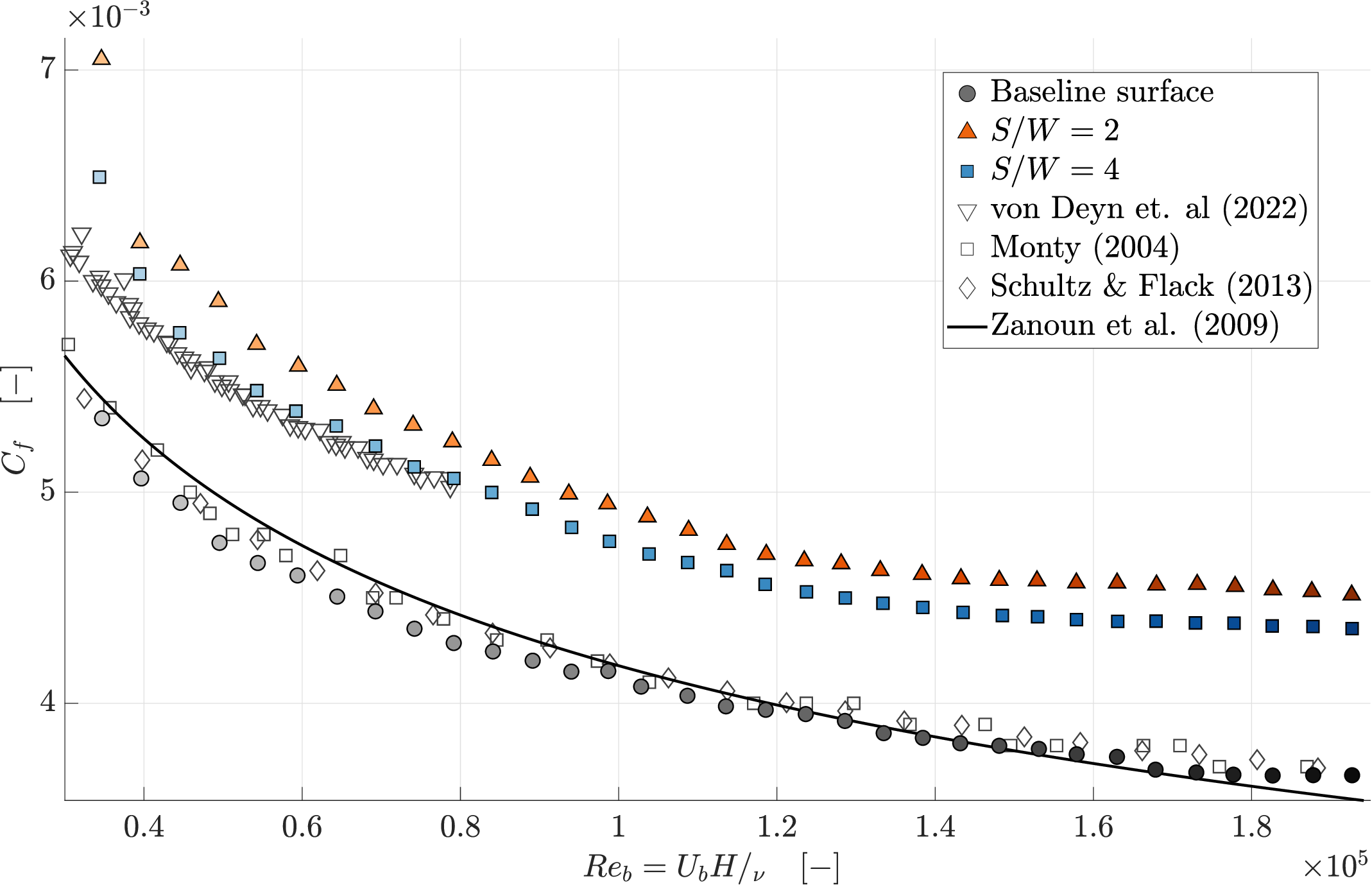}
    \caption{Skin friction coefficient for varying bulk Reynolds number measured for the two ridge surfaces. Filled grey markers denote the baseline smooth wall. Black diamond markers:  \citet{Schultz2013}, black square markers: \citet{Monty2005} and solid black line: \citet{Zanoun2009} included for reference. }
    \label{fig:ChannelRidgeCf}
\end{figure}

Figure \ref{fig:ChannelRidgeCf} presents the global skin friction coefficient $C_f$ as a function of bulk Reynolds number, $Re_b$, for both ridge surfaces and a baseline smooth wall. Reference data from \citet{Monty2005}, \citet{Zanoun2009} and \citet{Schultz2013} are included for smooth wall comparison to similar channel facilities. The ridge-type surfaces have a significant drag increase compared to the smooth wall surface, on the order of $15-20\%$, highlighting the drag penalty impact of the ridges. The skin friction coefficient of the $S/W = 4$ surface matches the results obtained by \citet{vonDeyn2021} (represented by downward-pointing triangles) for a similar ridge geometry ($S/W \approx 5$), although the drag is marginally higher in the Reynolds number range available ($Re_b \leq 80,000$). 

At higher Reynolds numbers (up to $Re_b \approx 170,000$), the rate of decrease in $C_f$ for the ridge surfaces diminishes (i.e. invariant within experimental uncertainty). In contrast, the skin friction coefficient of the smooth wall continues to gradually reduce, although slowly deviating from the empirical fit of \citet{Zanoun2009} similarily to the smooth wall surfaces of \citet{Monty2005} and \citet{Schultz2013} (possibly due to minor surface imperfections). This divergence between the smooth and ridge curves, particularly at higher $Re_b$, indicates a potential change in the underlying drag mechanism. To further explore this, we examine the roughness function $\Delta U^+$. This can be done by comparing the $C_f$ of a rough surface with the smooth wall at matched friction Reynolds number $Re_\tau$. The roughness function of the ridge-type surfaces is related to the difference in skin friction coefficients of the baseline and ridge-type surface through equation \ref{eq:dUpfromCf} \citep[see for example][]{Flack2019}

\begin{figure}
    \centering
    \includegraphics[width=0.4\textwidth]{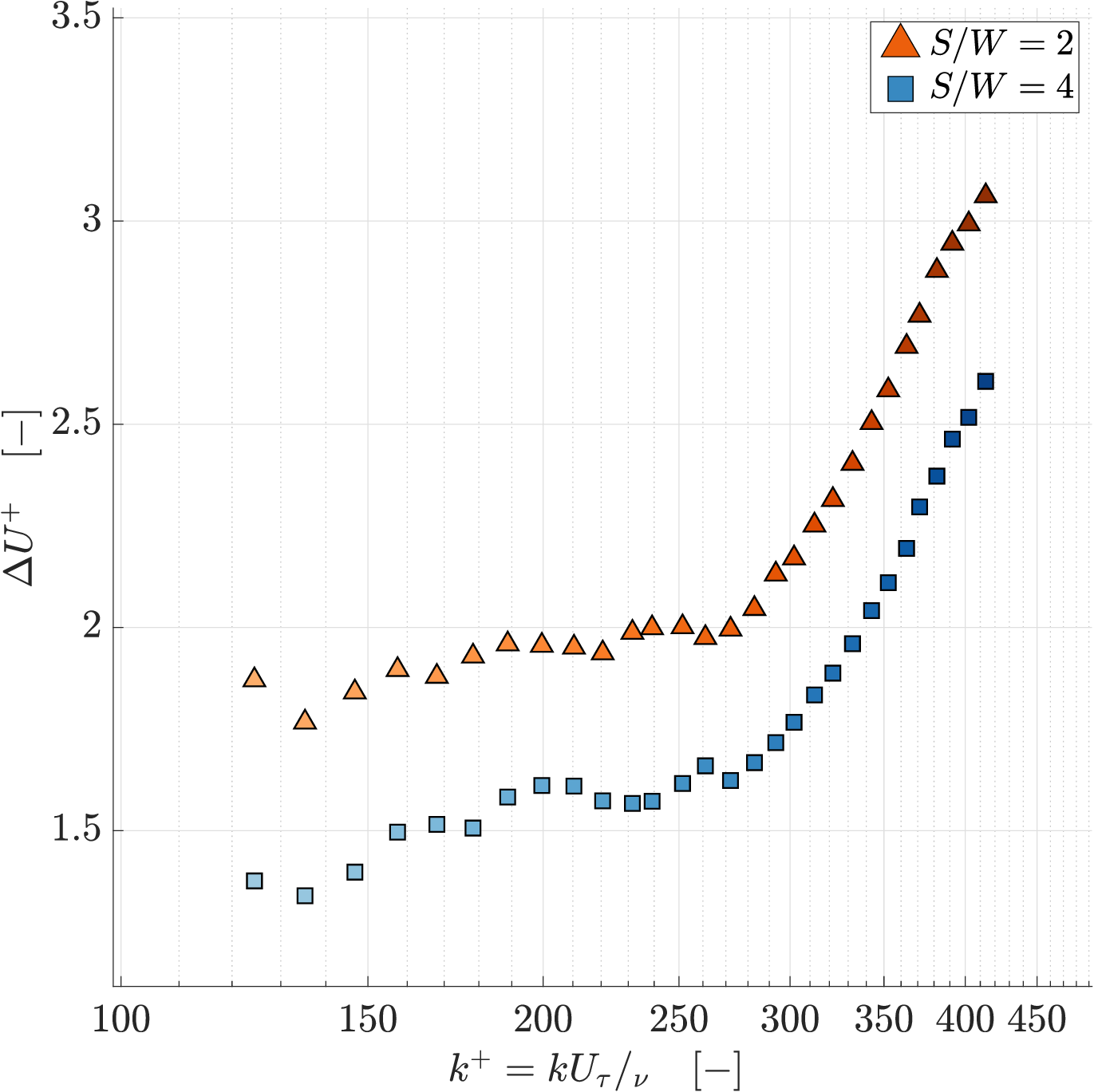}
     \includegraphics[width=0.59\textwidth]{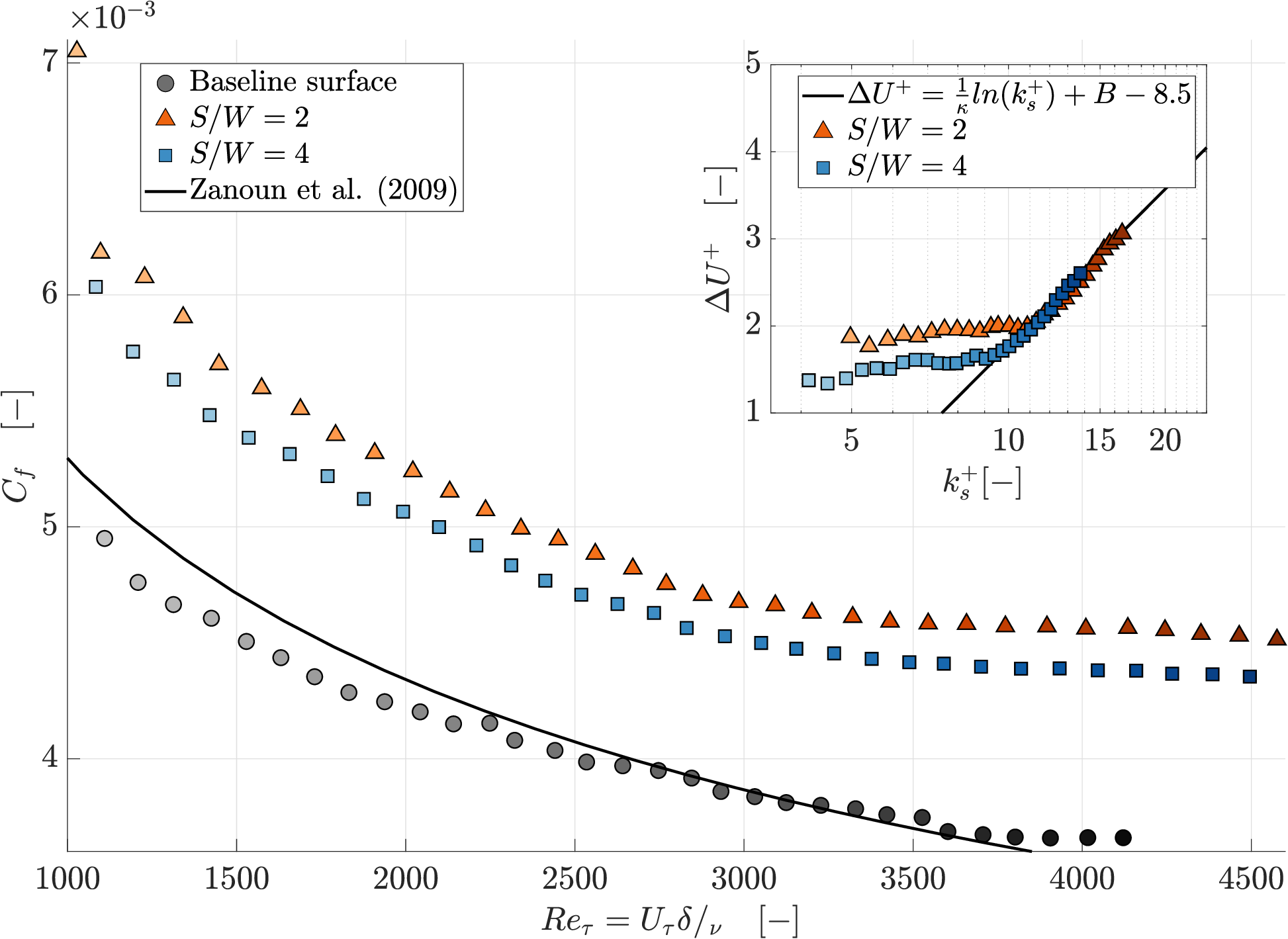}

    \caption{(a) The calculated roughness function as a function of the roughness Reynolds number $\Delta U^+ = f(k^+)$ based on $C_f$ data, (b) $C_f$ versus $Re_\tau$ that was used to calculate the roughness function. The inset in this figure is the collapse of the roughness function with $k_s$ determined by from the log-linear asymptote. }
    \label{fig:Channelkplus}
\end{figure}

\begin{equation}\label{eq:dUpfromCf}
    \Delta U^+ = \sqrt{\frac{2}{C_{f,s}}} - \sqrt{\frac{2}{C_{f,r}}}
\end{equation}

where $C_{f,r}$ is the ridge-wall skin friction coefficient and $C_{f,s}$ is the baseline (``smooth") wall skin friction coefficient. Note that this relationship has an inherent assumption of outer-layer similarity in the mean flow. This include the presence of the log region in the inner-layer where the roughness function can be measured (or attributed to) and an outer-wake that is similar between smooth and rough wall flows. This is not necessarily true for flows that have secondary motions (like the one used here) and will have to be separately confirmed. This is discussed further in the following sections. 
%




Figure \ref{fig:Channelkplus}(a) shows  the roughness function versus roughness Reynolds number $\Delta U^+ = f(k^+)$, where the ridge height $k$ was taken as the roughness length-scale. The roughness function was determined from the $C_f$ data shown in figure \ref{fig:Channelkplus}(b). At lower Reynolds numbers accessible in these experiments ($Re_\tau  \sim 1000$), the roughness function exhibits a very slow increase. Comparison of the two surfaces indicates that the gradual growth at lower $k^+$ could be different for different surfaces. This is consistent with observations in homogeneous rough surfaces where the transitionally-rough regime is not expected to be universal. For $k^+ > 300$, the roughness function of both surfaces appear to approach the log-linear asymptote as observed for typical fully-rough regime of a homogeneous surface. Once the data exhibits fully-rough regime, the roughness function can be related to the equivalent sand grain roughness:

\begin{equation}
    \Delta U^+(k_s^+) = (1/\kappa) ln(k_s^+) + B - B_s(k_s^+)
\end{equation}

where $B$ is the smooth wall log-law intercept and $B_s(k_s^+)$ is the log-law intercept for the ridge wall velocity profile ($U_R^+$). In the fully rough regime where $k_s^+ \rightarrow \infty$, $B_s(\infty) = 8.5$ as measured by \citet{Nikuradse1933}. Assuming for now that the ridge-type surfaces presented here do exhibit fully rough conditions, the equivalent sand grain roughness can be determined by shifting the log-scale x-axis by multiplying $k^+$ by a factor $C$ such that the log-linear asymptote in Figure \ref{fig:Channelkplus} collapses onto the fully rough asymptote of \citet{Nikuradse1933} \citep[for example, see][]{Chung2021}

\begin{equation}
    \Delta U^+ = (1/\kappa) ln(C k^+) + B - B_s(\infty) = (1/\kappa) ln(k_s^+) + B - B_s(\infty)
\end{equation}

where the smooth wall constants $\kappa = 0.384$, $B = 4.27$ from \citet{Lee2015} and $B_s(\infty) = 8.5$ were used \citep{Nikuradse1933}. The inset in \ref{fig:Channelkplus}(b) shows the collapse of the roughness function based with $k^+_s$. Although this collapse is reminiscent of fully rough behaviour, the very low values of $k_s^+$ (and $\Delta U^+$) is very surprising. This distinction is important since, unlike homogeneous rough walls where pressure drag dominates \citep{Flack2014}, the smooth ridges running the entire length of the channel here primarily induces only viscous stress. The absence of an obvious pressure drag mechanism for these smooth ridges suggests that alternative factors may be responsible for the observed skin friction behaviour, warranting further study. There are two plausible hypothesis that could explain this behaviour. First is that the secondary flows are Reynolds number dependent and the roughness function becomes full-rough like when as the strength of the secondary motions stops increasing with Reynolds number. This is unlikely as comparison of strength of secondary motions across different studies (with DNS, experiments etc), its strength does not appear to change with Reynolds number. This would need to be confirmed further with and will be discussed in the later section. The second and perhaps a more appealing hypothesis, is the following: as the instantaneous secondary motions meander in the spanwise direction, they impart normal flow on the side-faces of the ridge, resulting in something similar to a pressure drag. As Reynolds number increase, this pressure drag will increase until it becomes Reynolds number independent, which will lead to a presence of fully-rough like asymptote at a much lower of $k^+_s$. This would mean that if we have a higher ridge (i.e larger value of $k/\delta$, then, the onset of fully-rough like behaviour could be at an even lower $k^+_s$. At this point, this remains a hypothesis and we need further detailed measurements/analysis near the side-face of the ridge. This could be a potential way forward to future studies (especially numerical ones where this information can be readily accessible). 


\subsection{Mean flow}

\begin{figure}
    \centering
    \includegraphics[width=\textwidth]{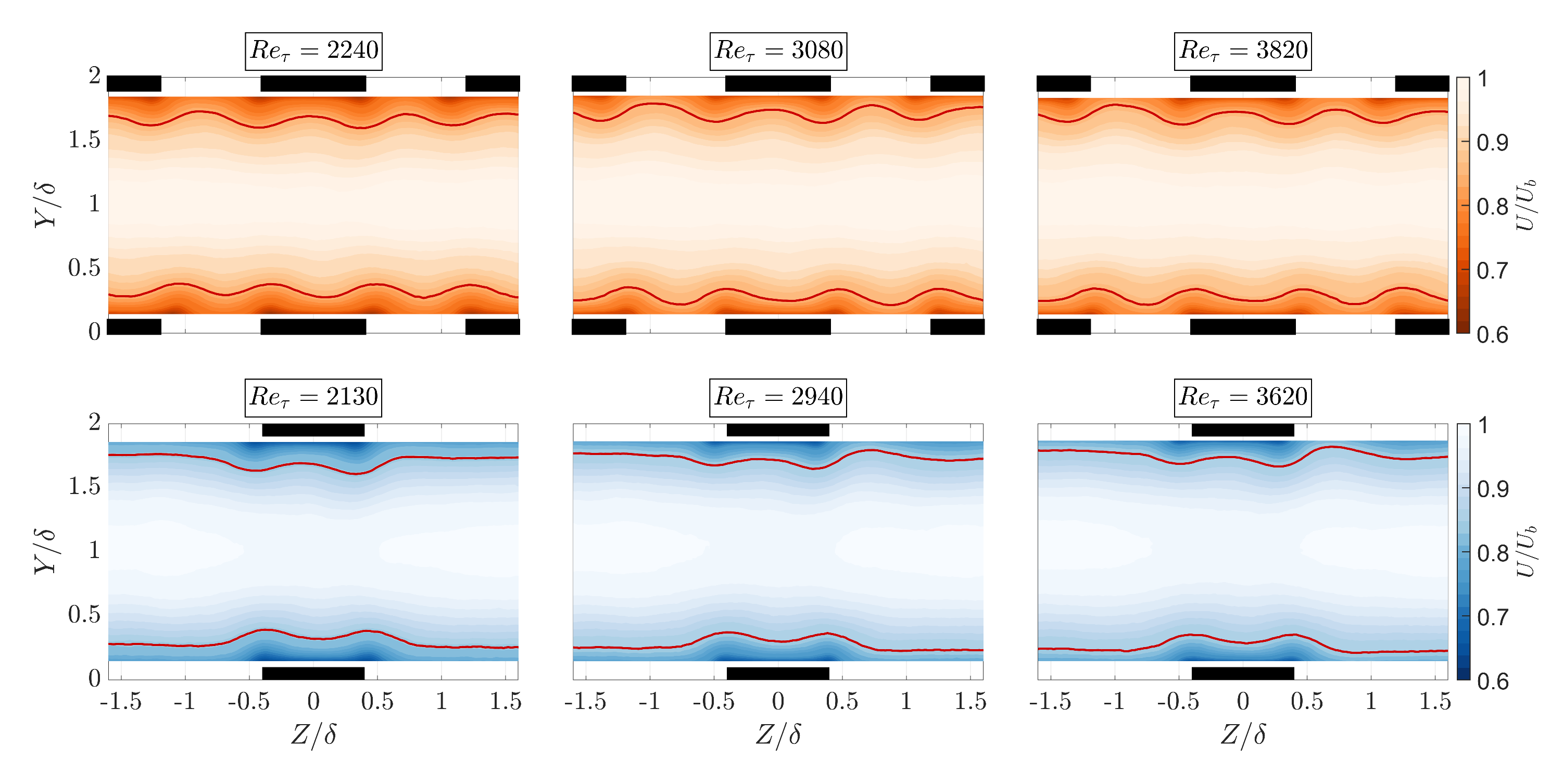}
    \caption[Modulation of the streamwise velocity components for both $S/W = 2$ and $S/W = 4$]{Modulation of the streamwise velocity components for both $S/W = 2$ (top row) and $S/W = 4$ (bottom row) at $Re_\tau \approx 2000, 3000, 4000$ (left to right). Red contour line highlights $U/U_b = 0.85$.}
    \label{fig:ChannelRidgeModulation}
\end{figure}

To investigate the Reynolds number independence of the secondary flows, cross-stream stereo PIV data is analysed.
Figure \ref{fig:ChannelRidgeModulation} shows the streamwise velocity contours at three different Reynolds numbers for both cases ($S/W$ = 2 and 4). The streamwise component clearly shows the formation of high- and low momentum pathways (HMPs and LMPs) for both cases with a noticeable spanwise variation near the ridge. A contour line at $U/U_b = 0.85$ highlights this modulation, which remains very similar across the Reynolds number range considered for both cases.

\begin{figure}
    \centering
    \includegraphics[width=\textwidth]{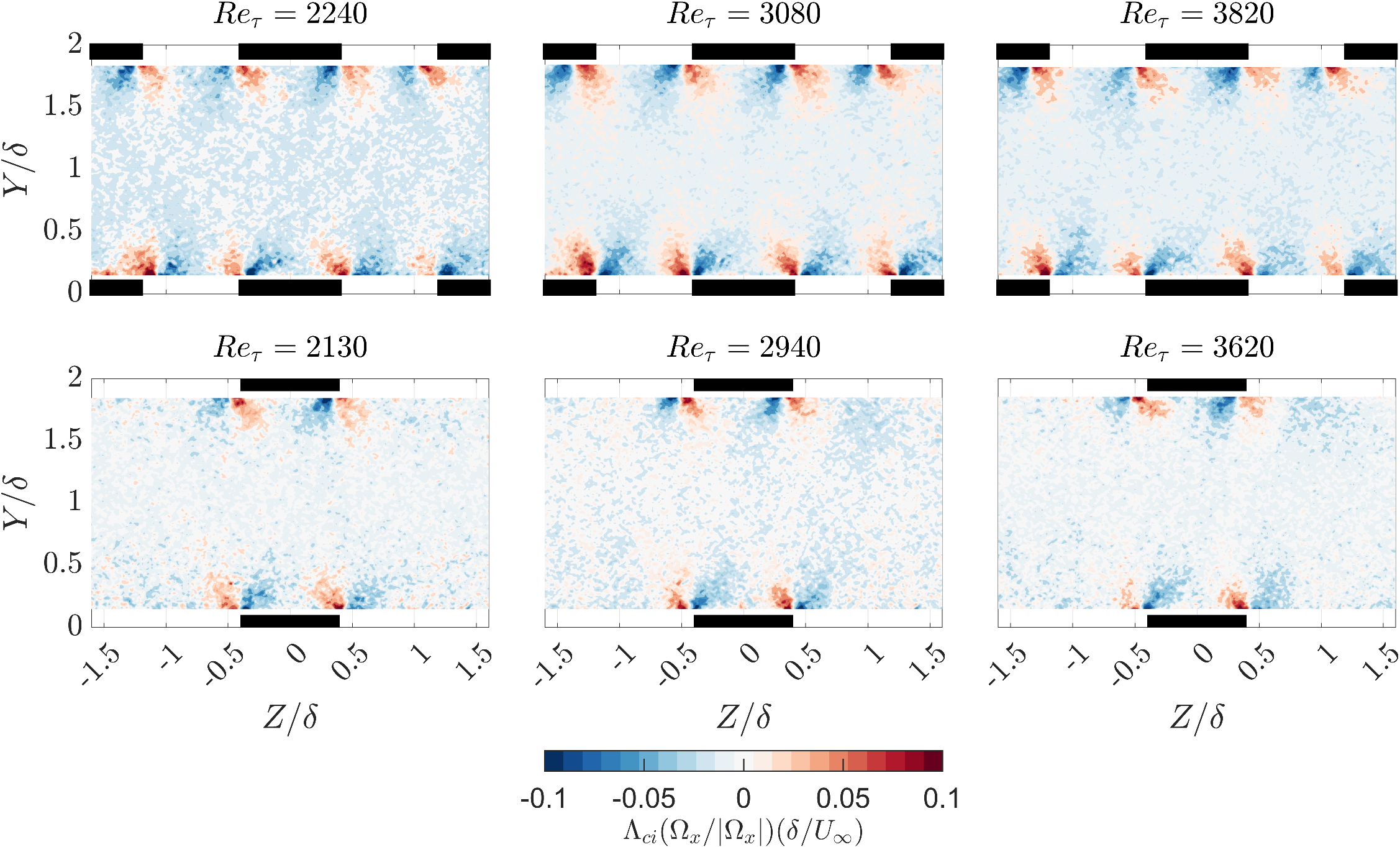}
    \caption[Mean swirling strength of the ridge cases.]{Mean swirling strength of the ridge cases. Top row: $S/W = 2$. Bottom row: $S/W = 4$. Columns correspond to increasing Reynolds number (left to right). Positive values denote clockwise rotation.}
    \label{fig:channel_SwirlStrength}
\end{figure}

The manifestation of observed secondary motion in other components is examined through the mean swirling strength, $\lambda_{ci}$,  which s computed as the imaginary part of the complex eigenvalue of the mean in-plane velocity gradient tensor \citep[see][for details]{Zhou1999}. Because the swirling strength itself lacks directional information it is multiplied by the sign of the mean streamwsie vorticity , $\Omega_x/|\Omega_x|$, where $\Omega_x = \partial V/\partial z - \partial W/ \partial y$, yielding the signed swirling strength. This value is then normalised by $\delta/U_\infty$ to produce the mean swirling strength shown in Figure \ref{fig:channel_SwirlStrength}.

A pair of counter-rotating vortices appear at the edges of the ridges ---- a common feature for wide rectangular ridges (see e.g. case H6 in \cite{Medjnoun2020}). Notably, the swirling strength is higher for the $S/W = 2$ case, supporting  the presence of stronger vortical structures that extract energy from the mean flow and contribute to increased drag. For both ridge configurations, the swirling strength and spatial extent of these structures remain largely unchanged across the Reynolds numbers considered here further reinforcing that the large-scale secondary motions are independent of Reynolds number. 

\begin{figure}
    \centering
        \includegraphics[width=0.48\textwidth]{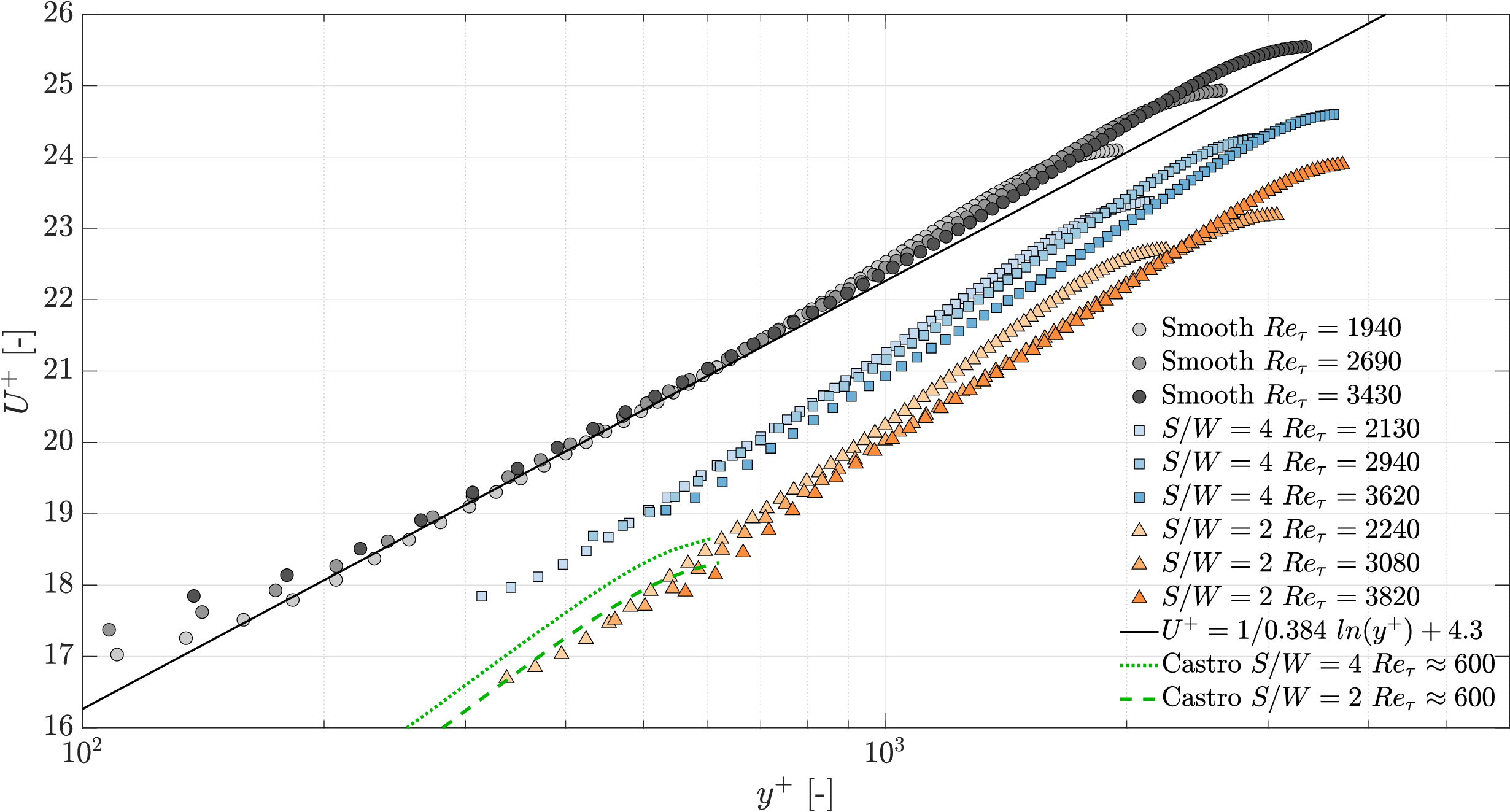}
        \includegraphics[width=0.48\textwidth]{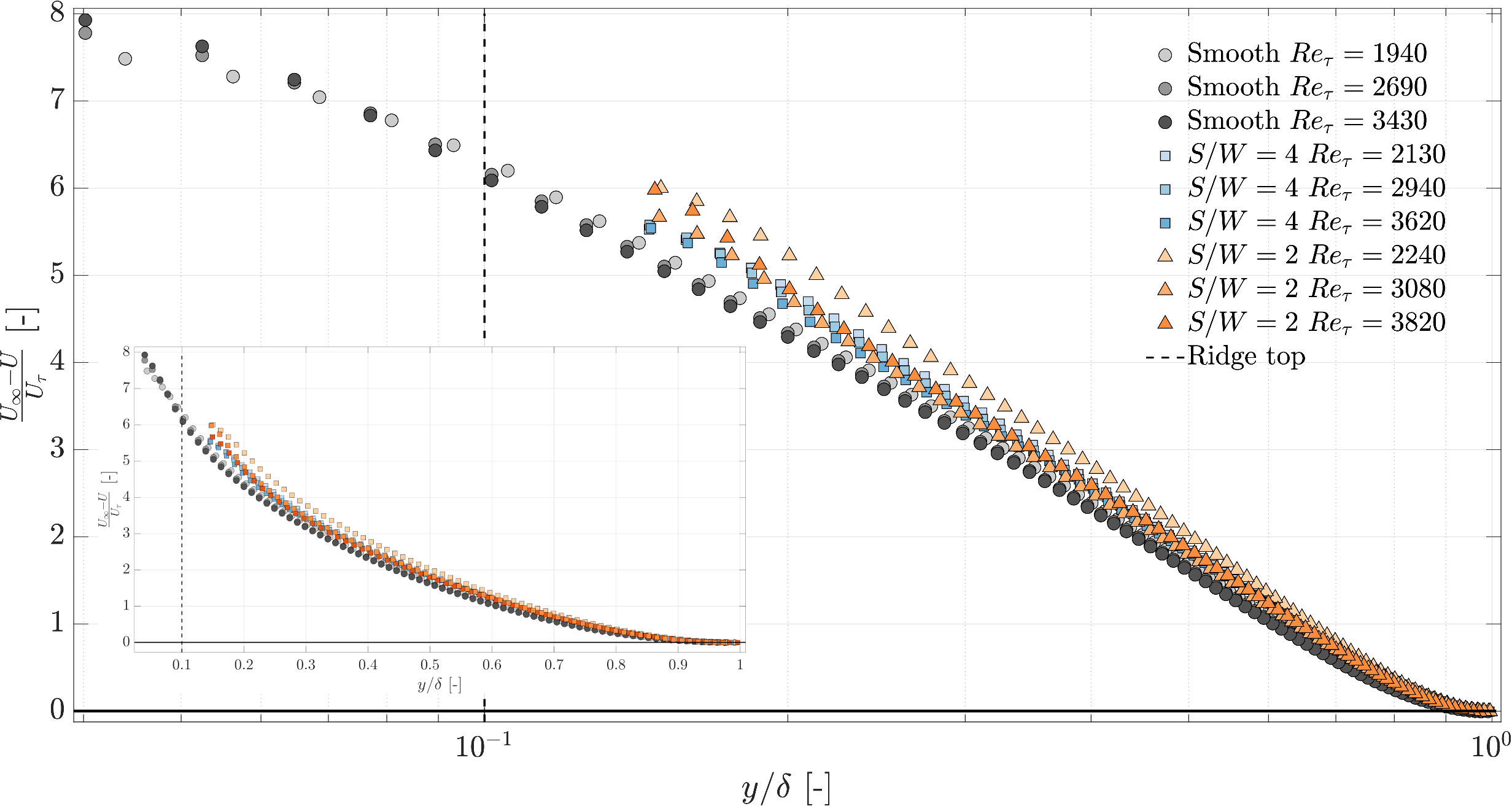}
        \includegraphics[width=0.6\textwidth]{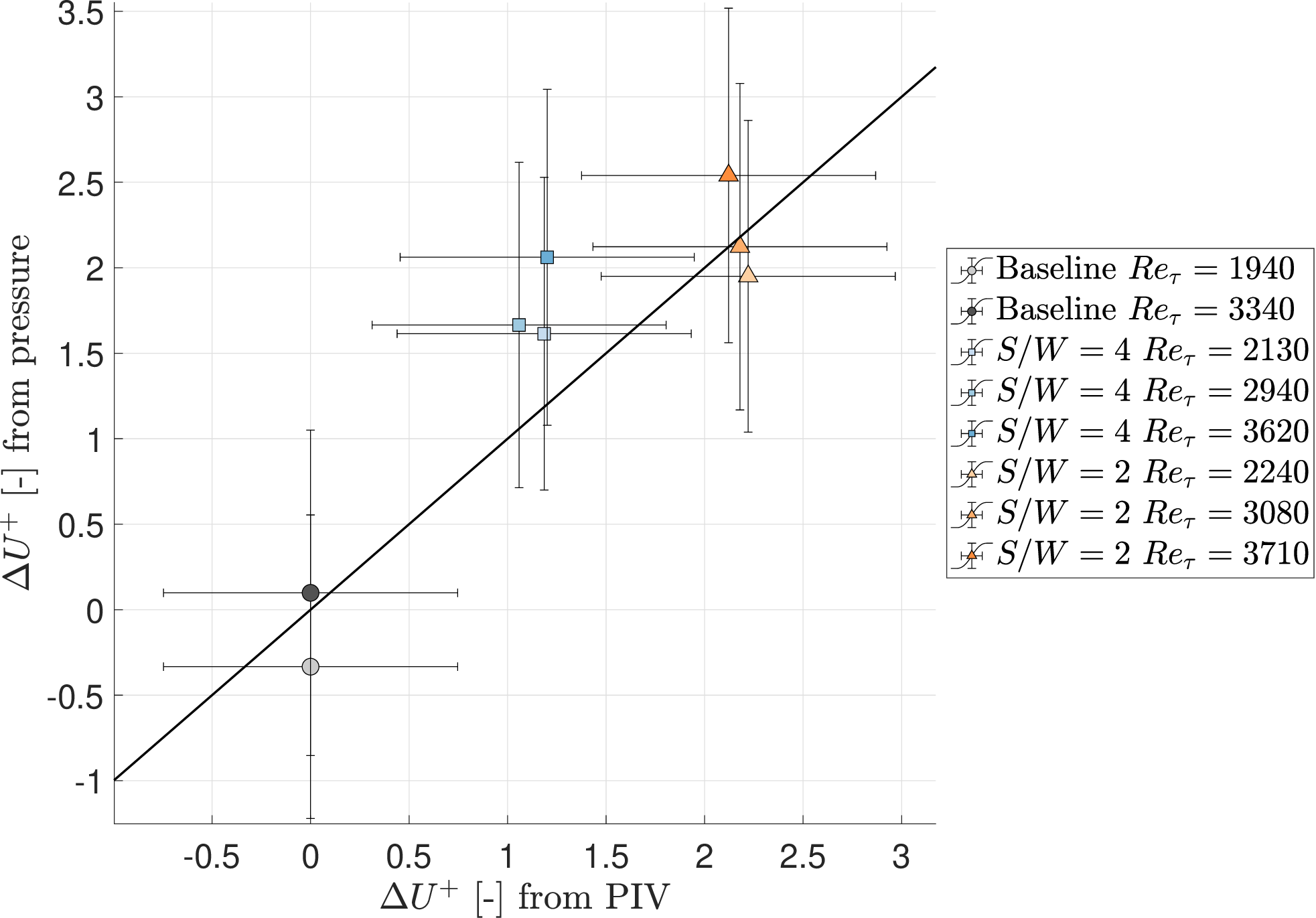}
    \caption{(a) Inner scaled, spanwise-averaged streamwise velocity profiles and (b) Spanwise-averaged streamwise velocity defect profiles.  The black dashed line indicates the ridge top. The inset shows the same data on a linear x-axis scale. For both figures: Grey markers: Baseline smooth-wall surface. Blue squares: $S/W = 4$. Orange triangles: $S/W = 2$. (c) Comparison between $\Delta U^+$ calculated from PIV velocity profiles and skin-friction measurements.}
    \label{fig:Channelprofiles}
\end{figure}

A more quantitative analysis of the mean flow can be carried out by examining the spanwise-averaged (over one wavelength) profiles of the mean streamwise velocity. Figures \ref{fig:Channelprofiles} a and \ref{fig:Channelprofiles} b show the inner-scaled mean velocity in logarithmic linear and defect forms, respectively. For comparison, profiles for the baseline smooth wall surface at the same $Re_\tau$ are also included (Note that data for the ridge cases is unavailable near and below the ridge top).  Figure \ref{fig:Channelprofiles}a shows that the ridge profiles are shifted downwards consistent with increasing skin friction, reflecting the variation in $\Delta U^+$ observed in Figure \ref{fig:Channelprofiles}b. The figure also includes a log-law fit with parameters from \citet{Lee2015} and DNS data from \citet{Castro2021} for similar ridge configuration at $Re_\tau \approx 600$. The DNS results agree well with the $S/W = 2$ case, but less so with the $S/W = 4$ case.  

The defect profiles in Figure \ref{fig:Channelprofiles}b do not collapse up to about $y/\delta = 0.6\sim0.8$ where the ridge surfaces exhibit a larger velocity defect, compared to smooth-wall indicative of momentum loss to sustain the secondary flows. Beyond this region the profiles collapse, suggesting reduced influence of the secondary flows. We also note that the lack of collapse between 0.2$\delta$ up to 0.6$\delta$ could also be due to incorrect virtual origin (or zero-plane displacement). It might be possible to collapse the profiles down to much lower distance from the wall if we account for this displacement. In these profiles, we used $h/2$ as our origin for the profiles. Recently, \cite{Chen2023} identified zero-plane displacement by collapsing the diagnostic function ($y^+dU^+/dy^+$) in the entire outer layer. Given the presence of secondary flows in the current study, this is perhaps not advisable. Regardless, we observe that  thecollapse is better closer to wall with increasing Reynolds numbers. This suggests that the estimate of $\Delta U^+$ from the $C_f$ curves as done in the previous section is likely to be more accurate for higher Reynolds numbers where there is a clear log region and the outer flow exhibits similarity with the smooth wall.

Figure \ref{fig:Channelprofiles}c compares $\Delta U^+$ obtained from the PIV velocity profiles with those derived from the skin-friction measurements. While the ordering of the cases is as expected from $\Delta U^+$, the uncertainty is fairly high --- up to approximately $\pm 1$ --- due to measurement uncertainty propagation. This is especially clear when looking at the values of $\Delta U^+$ of the baseline smooth-wall that shows minor variations around zero indicating the difficulty in determining the roughness function with this level of accuracy from experimental data. We also note that uncertainty is sensitive to the choice of $\kappa$ and the selected log region (although this was done consistently across all cases). Therefore, caution must be exercised while interpreting $\Delta U^+$ from PIV data, especially when the roughness function is relatively small. 

Overall, mean flow analysis demonstrates that while the mean flow modulation varies with surface condition, its overall structure remains largely unchanged with increasing Reynolds number. However, it is unclear if this invariance is limited to mean flow or extends to turbulent statistics and structure as well and this is explored in the next section. 

\subsection{Turbulence statistics}

\begin{figure}
    \centering
  \includegraphics[width=0.8\textwidth]{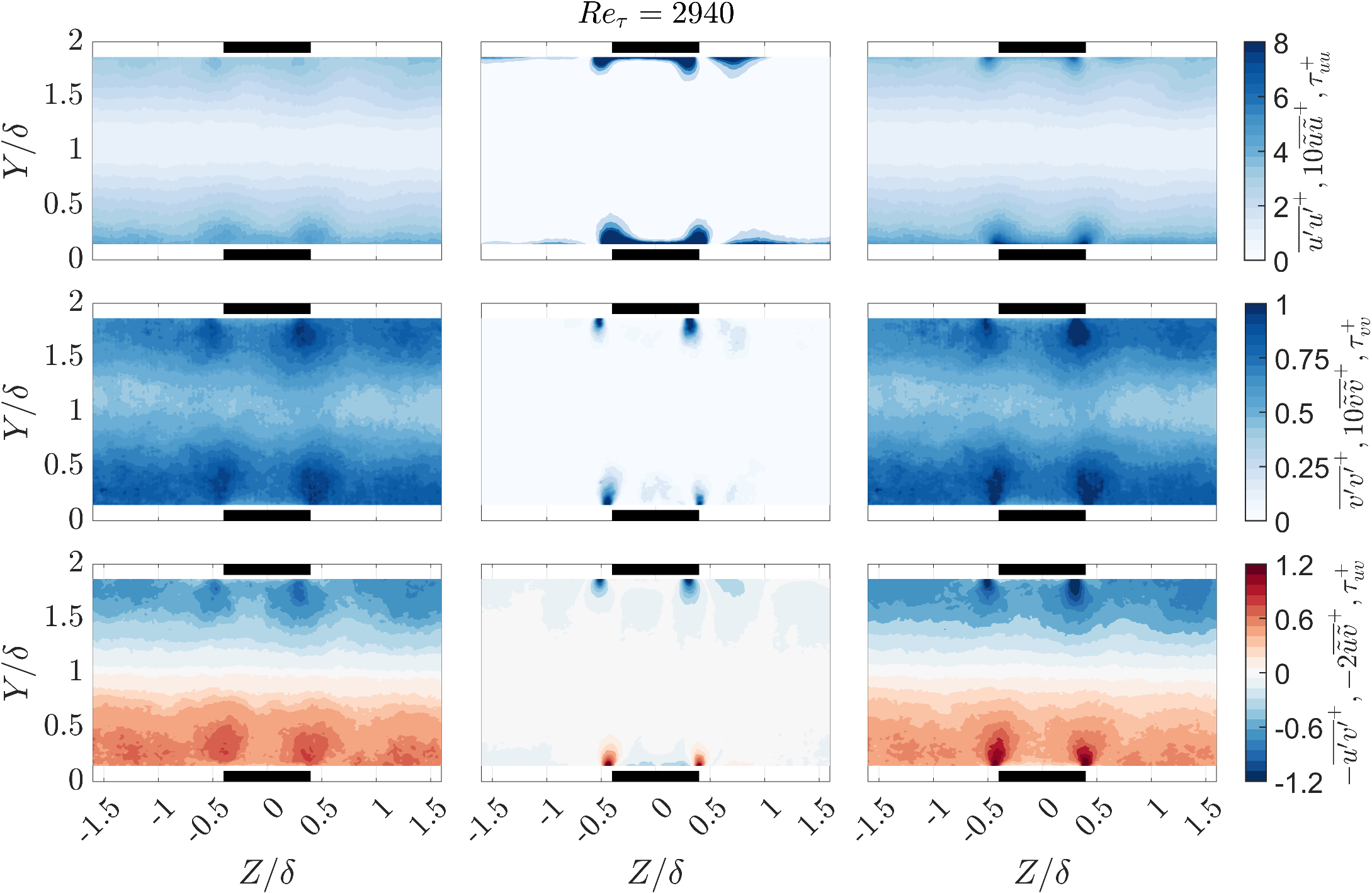}
   \includegraphics[width=0.8\textwidth]{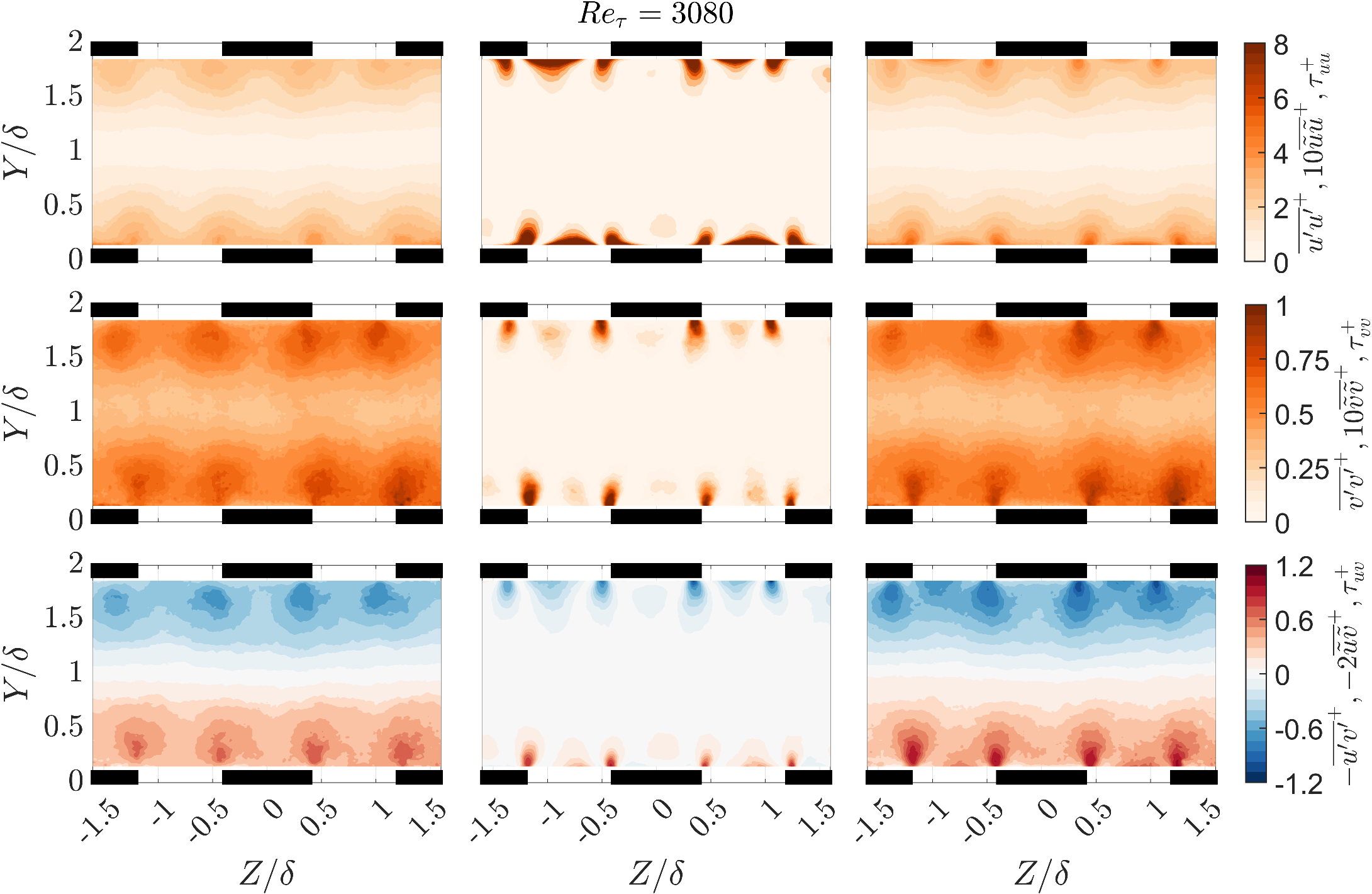}
\caption[Contours of inner scaled stress components at the intermediate Reynolds number $Re_\tau \approx 3000$ for the $S/W = 4$ surface (top) and $S/W = 2$ (bottom).]{Contours of inner scaled stress components at the intermediate Reynolds number $Re_\tau \approx 3000$ for the $S/W = 4$ surface (top) and $S/W = 2$ (bottom). For both figures, from left to right: Reynolds stresses, dispersive stresses and approximated total stress. From top to bottom: $uu$, $vv$ and $uv$ components. Dispersive stresses are multiplied by a factor $10$ (for $\overline{\tilde{u}\tilde{u}}^+$ and $\overline{\tilde{v}\tilde{v}}^+$) and $2$ (for $\overline{\tilde{u}\tilde{v}}^+$) for better visualisation.}
    \label{fig:channel_stressconts}
\end{figure}

Figure \ref{fig:channel_stressconts} shows contours of inner-scaled stress components for the larger spacing case $S/W = 4$ (top) and $S/W = 2$ (bottom) at $Re_\tau \approx 3000$. The left column shows the Reynolds stress components $\overline{u'u'}^+$, $\overline{v'v'}^+$ and $-\overline{u'v'}^+$ (from top to bottom). The secondary flows generate stress concentrations at the edges of the ridges for all three components. In the middle column, the dispersive stress components $\overline{\tilde{u}\tilde{u}}^+$, $\overline{\tilde{v}\tilde{v}}^+$ and $-\overline{\tilde{u}\tilde{v}}^+$ are shown. Nonzero dispersive stresses directly indicate mean flow heterogeneity, in this case induced by ridge-generated secondary flows. Because their magnitudes are typically much smaller than those of the Reynolds stresses, the dispersive stresses are displayed here multiplied by factors of 10 (for the $\overline{\tilde{u}\tilde{u}}^+$ and $\overline{\tilde{v}\tilde{v}}^+$ components) and 2 (for the $-\overline{\tilde{u}\tilde{v}}^+$) to facilitate visualisation. The total stress, $\tau$, in the flow comprises the Reynolds stress, the dispersive stress and this is shown in the right column in Figure \ref{fig:channel_stressconts}. The $S/W = 2$ case exhibits stronger local stress magnitudes when compared to $S/W$ = 4, in the Reynolds, dispersive and total stress fields. This indicates enhanced secondary flow intensity, which is consistent with the results of the linear model used to design the experiments. For both surfaces, the secondary flows of the top and bottom walls do not interact in a time-averaged sense; the turbulent statistics are confined to approximately $y/\delta \approx 0.6$, leaving the channel core with low turbulence levels.

It should be noted that due to resolution limitation of the PIV measurements, the dispersive stresses near the wall and immediately above the ridge centers (where spanwise fluctuations are low) are under-resolved. In particular, reflections above the ridge centers results in low signal-to-noise ratios with increased uncertainty. Thus, the expected signatures of weak tertiary flows are not clearly observed. This is particularly important when we compare the data across different Reynolds numbers and this is further discussed below. 

\begin{figure}
    \centering
    \includegraphics[width=0.8\textwidth]{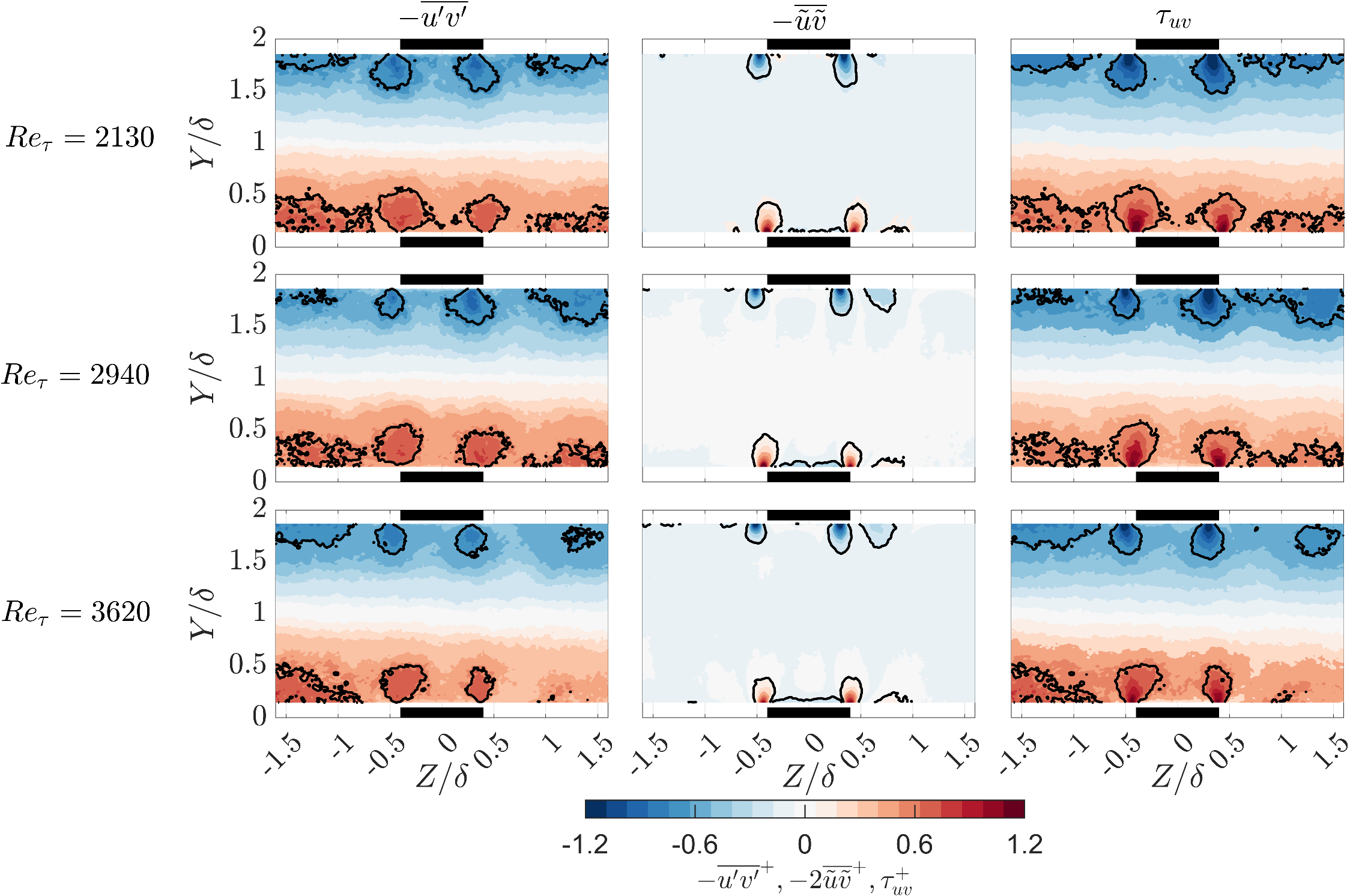}
        \includegraphics[width=0.8\textwidth]{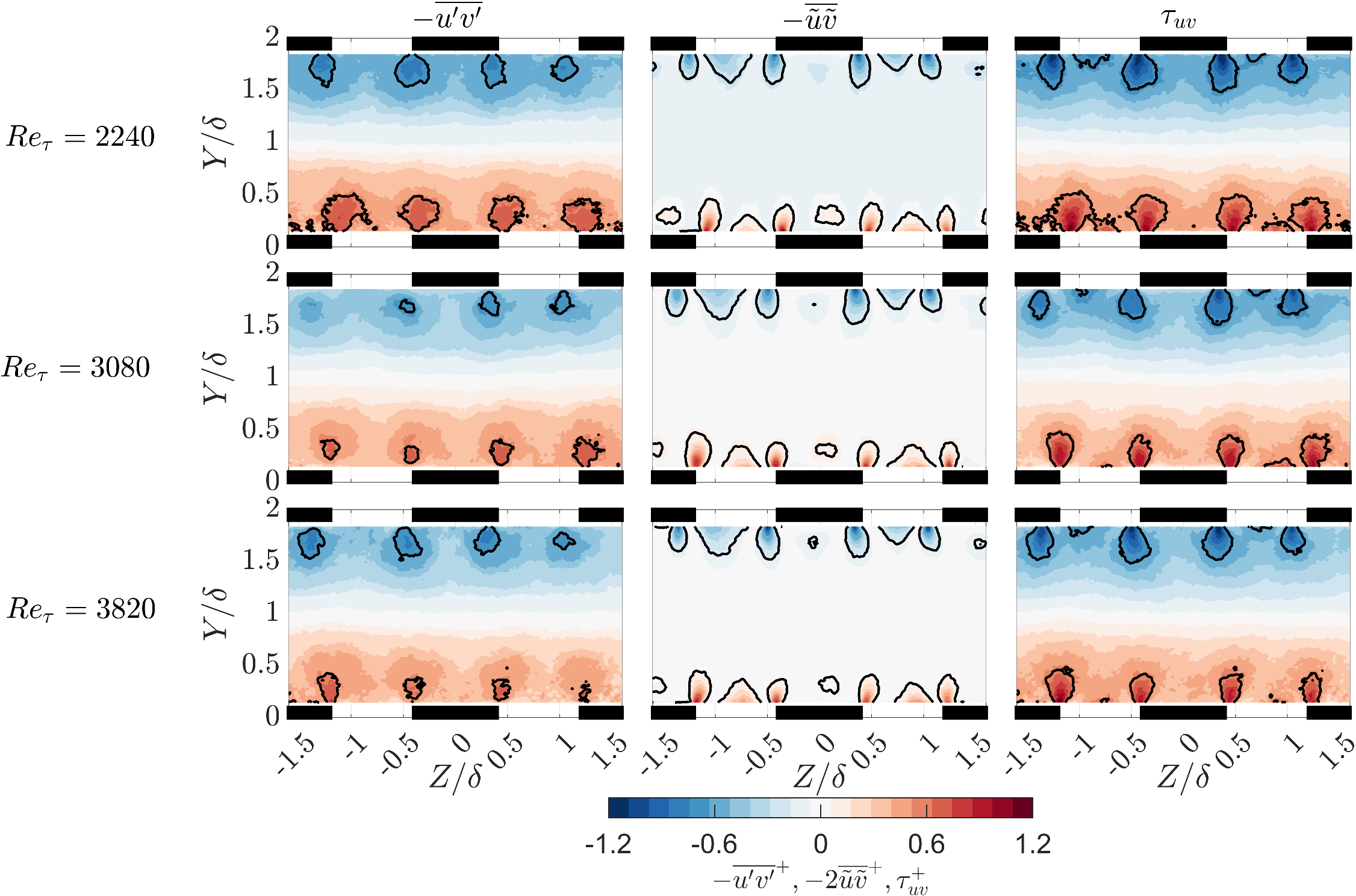}
    \caption[Contours of inner scaled shear stress components ($uv$) of the three stress tensors at two Reynolds numbers for the $S/W = 4$  and $S/W = 2$.]{Contours of inner scaled shear stress components ($uv$) of the three stress tensors at two Reynolds numbers for the $S/W = 4$  and $S/W = 2$. From left to right: Reynolds shear stress, dispersive shear stress, and approximated total shear stress. Dispersive stresses are multiplied by 2 for clarity. The black solid line indicates contour levels at $\pm0.6$ (or $\pm0.06$ for the dispersive component).}
    \label{fig:channel_ReVarShearconts}
\end{figure}

Figure \ref{fig:channel_ReVarShearconts} shows the inner-scaled turbulent, dispersive and total shear stress components  ($uv$) for the $S/W = 4$ (top) and $S/W = 2$ (bottom) case at all two Reynolds numbers. A contour levels of $\pm0.6$ (or $\pm0.06$ for the dispersive component) is overlaid to highlight the secondary flow features. It is observed that both the spatial extent and magnitude of these stresses are largely unaffected by increasing Reynolds number. This confirms the hypothesis from the previous section that the strength of the large-scale secondary motions are Reynolds number. We note that the spatial resolution of the PIV measurements is poorer with increasing Reynolds number, however, the stresses are computed with filtered data where the spatial resolution of the different Reynolds numbers are matched (in inner units).

\begin{figure}
    \centering
    \includegraphics[width=\textwidth]{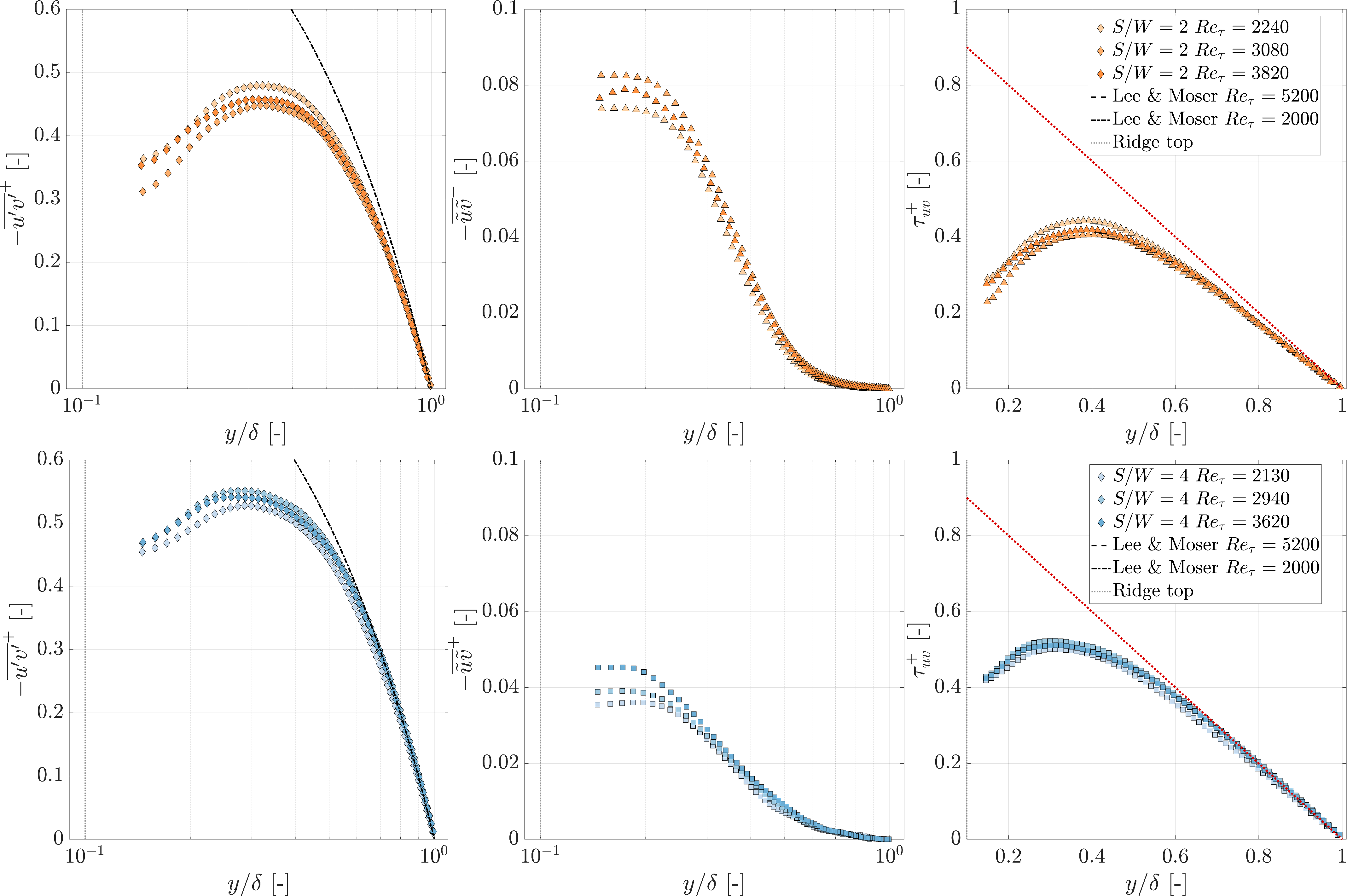}
    \caption[Shear stresses in the channel for both surfaces.]{Shear stress profiles in the channel for both surfaces. Left: Reynolds shear stress component $-\overline{u'v'}^+$. (Black dashed and dash-dotted lines correspond to smooth wall DNS by \cite{Lee2015} at $Re_\tau = 5200$ and $Re_\tau = 2000$) Middle: Dispersive shear stress component $-\overline{\tilde{u}\tilde{v}}^+$. Right: Approximate total shear stress $\tau_{uv} \approx \overline{u'v'}^+ + \overline{\tilde{u}\tilde{v}}^+$. The red dashed line indicates the theoretical total shear stress distribution.}
    \label{fig:channel_uvStressGrid}
\end{figure}

Figure \ref{fig:channel_uvStressGrid} shows inner-scaled, spanwise-averaged (over one wavelength) Reynolds shear stress ($uv$) profiles for both ridge surfaces. Turbulent, dispersive and total stresses are shown in the three columns. All three stresses show collapse in the outer region (beyond $y/\delta \approx 0.6$). Although the limited spatial resolution leads to an underestimation of stress magnitudes near the wall, however, the filtered data allows for comparison at matched resolution across Reynolds number. It is observed that shear stress component for $S/W = 4$ surface has a larger peak than the $S/W = 2$ surface across all Reynolds numbers. This behaviour is consistent with the expectation that increased drag (and hence higher $C_f$) in the $S/W = 2$ case may be partially attributed to an increased contribution from ridge side friction and near-wall viscous effects. Furthermore, the peak shear stress in the $S/W = 2$ profiles occurs slightly further away from the wall ($y/\delta \approx 0.35$) than for the $S/W = 4$ case ($y/\delta \approx 0.25$), suggesting that the secondary flows in the denser ridge configuration affect a larger portion of the outer layer. Furthermore, for the $S/W = 2$ surface (top panels), the dispersive shear stress is significantly higher near the ridge top --- roughly by a factor of two --- indicating a stronger spanwise heterogeneity in the mean flow. Notably, the peak in dispersive stress coincides with a dip in the Reynolds shear stress. When these components are summed, the $S/W = 4$ surface agrees better with the theoretical total shear stress distribution for smooth-wall channel flow (red dashed line), suggesting that a larger portion of the outer flow is nearly homogeneous. Importantly, for both ridge configurations, the turbulent statistics exhibit minimal variation with Reynolds number over the range considered. 

\subsection{Turbulent structures}

 \begin{figure}
    \centering
    \includegraphics[width=\textwidth]{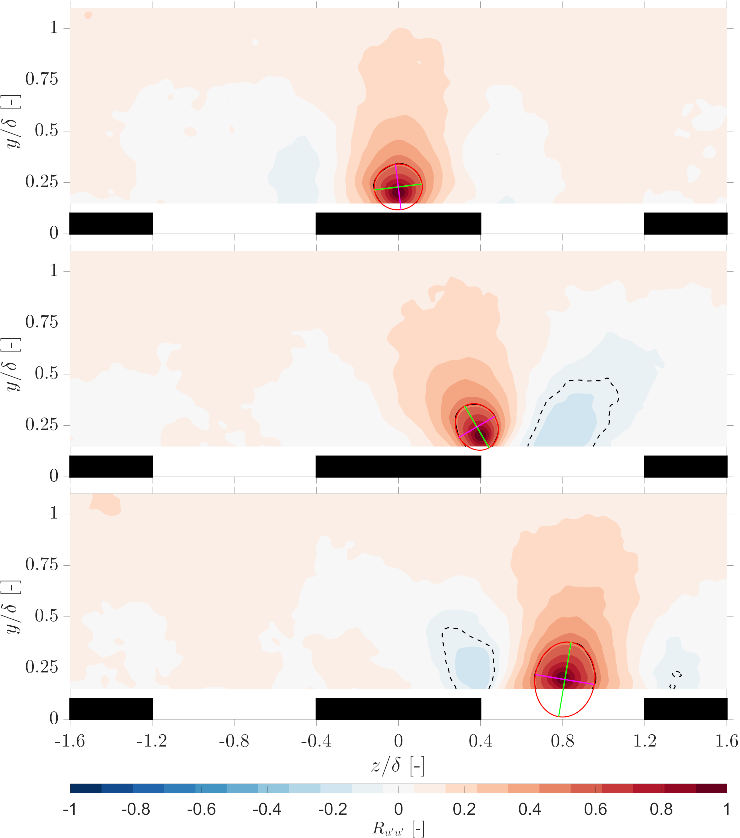}
    \caption{Sample two-point correlation contours of $R_{\overline{u'u'}}$ for the $S/W = 2$ case at $Re_\tau \approx 3000$. Top: reference point $(z_{ref}, y_{ref}) = (0\delta, 0.2\delta)$ (above the ridge center). Middle: $(z_{ref}, y_{ref}) = (0.4\delta, 0.2\delta)$ (above the ridge edge). Bottom: $(z_{ref}, y_{ref}) = (0.8\delta, 0.2\delta)$ (above the valley). The solid black line shows the $0.5$ contour, and the dashed line shows $-0.1$. The fitted ellipsoid (red) with its major (green) and minor (magenta) axes is also displayed.}
    \label{fig:channel_2ptcorrcont}
\end{figure}

We calculate a two-point correlations of velocity fluctuations in the spanwise--wall-normal plane as, 

\begin{equation}\label{eq:twoptcorr}
    R_{u_i'u_j'}(z_{ref}, y_{ref}) = \frac{\overline{u_i'(z_{ref}, y_{ref})u_j'(z, y)}}{\sigma_{u_i'}(z_{ref}, y_{ref})\sigma_{u_j'(z, y)}}
\end{equation}

where $(z_{ref}, y_{ref})$ is the reference point at which the correlation of the fluctuating velocity component $u_i'$ with the component $u_j'$ at any point $(z, y)$ is computed. Here, $\sigma_{u_i'}$ denotes the standard deviation of $u_i'$. Previous work has demonstrated significant spatial variation on the coherence of turbulent structures by examining the $R_{u'u'}$ component \citep[see for example][]{Kevin2019a, Medjnoun2020}. In the present study, we extend that investigation to cover different Reynolds numbers to examine the influence of Reynolds number variations on the turbulent structures. The two-point correlation is computed on a grid spanning the entire half-channel height (assuming symmetry between the top and bottom halves) in the wall-normal ($y$) direction and one wavelength in the spanwise ($z$) direction.


Contours for $R_{u'u'}$ is shown in Figure \ref{fig:channel_2ptcorrcont} for the $S/W = 2$ case at $Re_\tau \approx 3000$. Three reference locations ($z_{ref},y_{ref}$) are illustrated: above the ridge center $(0, 0.2\delta)$ in the top panel, above the ridge edge ($(0.4\delta, 0.2\delta)$) in the middle panel, and above the valley ($0.8\delta, 0.2\delta)$) in the bottom panel. In these plots, the solid black line represents the $R_{u'u'} = 0.5$ contour, while the dashed black line indicates $R_{u'u'} = -0.1$. Typically, the negative correlation appears when a high-momentum region (HMP, where $u'>0$) is compared with a low-momentum region (LMP, where $u'<0$), and vice versa. For each reference points, we fit an ellipsoid to the $R_{u'u'} = 0.5$ contour (red line in Figure \ref{fig:channel_2ptcorrcont}). The major and minor axes of the ellipsoid --- shown as green and magenta lines, respectively --- provide quantitative measures of the spatial coherence of the turbulent structure.

For both ridge configurations, the HMP over the ridge (associated with a downwash region) exhibits a large area of positive correlation extending through most of the boundary layer, although the $R_{\overline{u'u'}}=0.5$ contour here is nearly circular, making the ellipsoid orientation ambiguous. At the LMP regions near the ridge edges, the correlation becomes weakly negative. At these locations, the fitted ellipsoid tilts toward the direction of the radial motions (i.e., perpendicular to the primary streamwise flow). In the valley region, the area enclosed by the ellipsoid is larger —-- stretching nearly to the wall --— which may be attributed to a larger local distance from the wall that imposes less constraint on the flow.

To quantify these observations, we define a characteristic length scale $L_{uu}$ as

 \begin{equation}
     L_{uu} = \sqrt{l_{maj}^2 + l_{min}^2}
 \end{equation}

 where $l_{maj}$ and $l_{min}$ are the lengths of the major and minor axes, respectively.

 \begin{figure}
    \centering
    \includegraphics[,clip,width=0.48\textwidth]{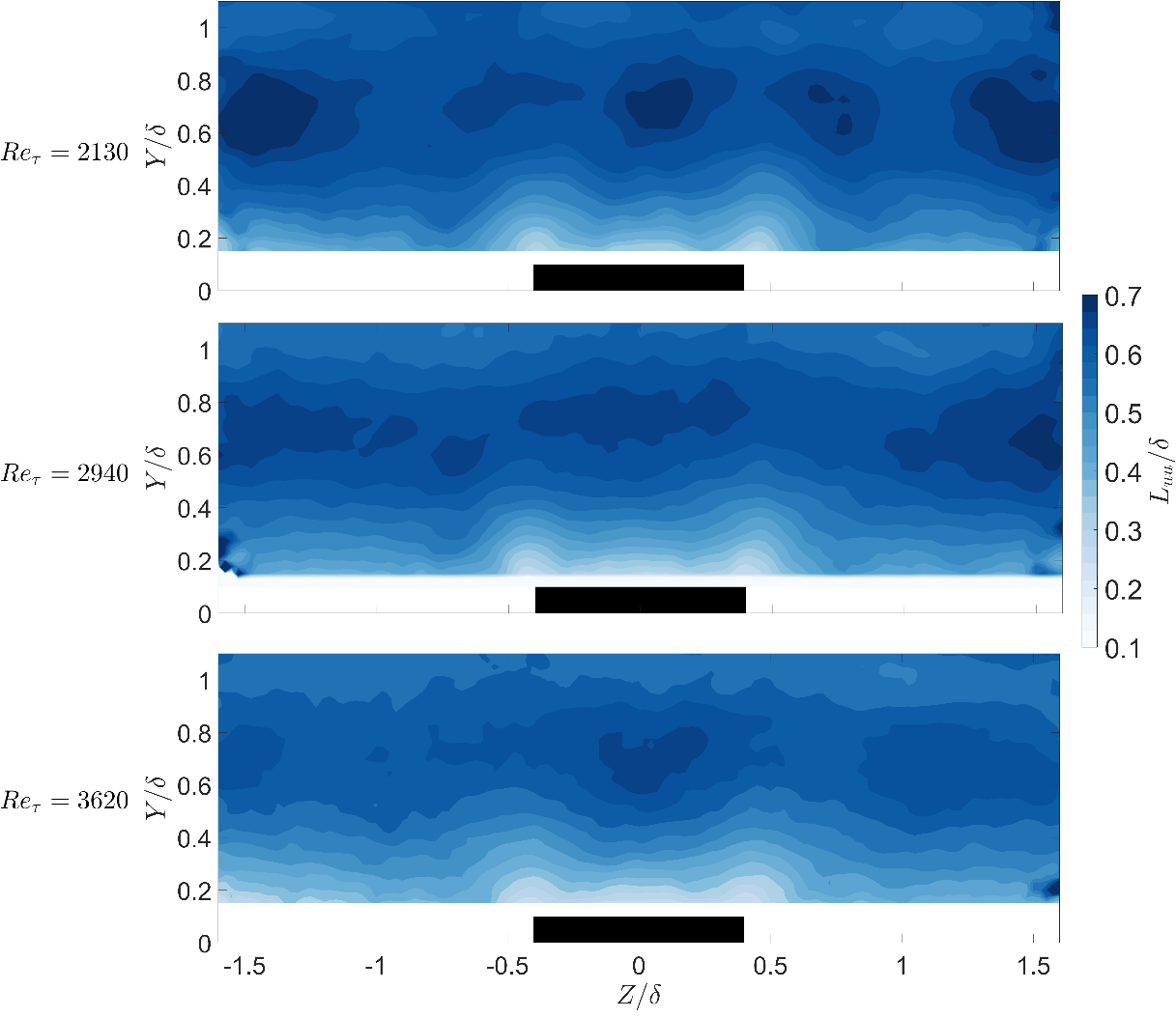}
     \includegraphics[width=0.48\textwidth]{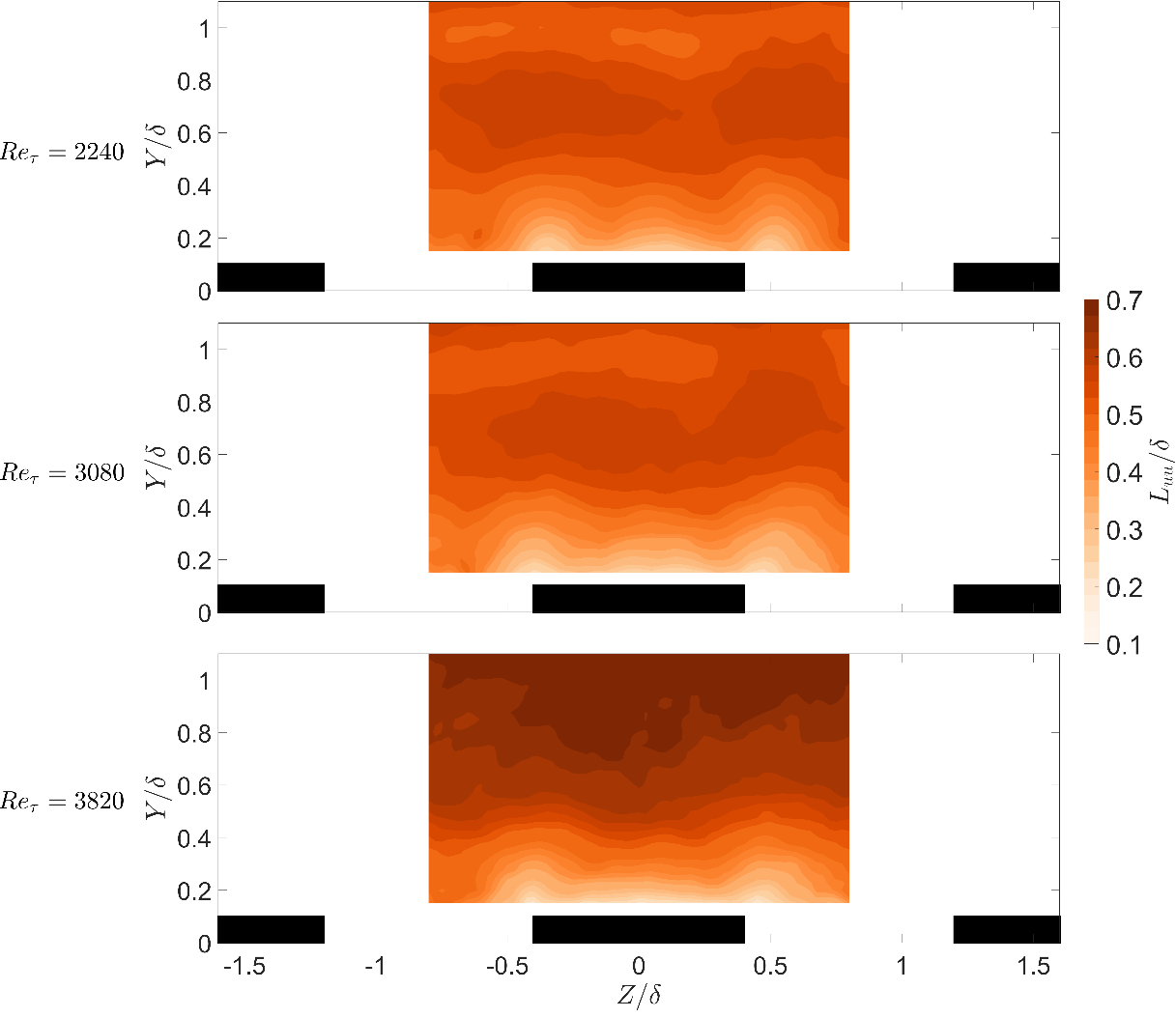}
    \caption[Characteristic correlation length for the $S/W = 4$ and $S/W = 2$ surface.]{Characteristic correlation length for the $S/W = 4$ and $S/W = 2$. Top: $Re_\tau \approx 2000$. Middle: $Re_\tau \approx 3000$. Bottom: $Re_\tau \approx 4000$}
    \label{fig:channel_corrlength}
\end{figure}

 Figure  \ref{fig:channel_corrlength} present contour maps of $L_{uu}$ for the $S/W = 2$ and $S/W = 4$ surfaces, respectively. For the $S/W = 2$ case, the correlation length is smallest near the wall in both HMP and LMP regions. Further from the wall, $L_{uu}$ increases and exhibits a pronounced spanwise variation: regions of high momentum display larger $L_{uu}$, whereas low-momentum regions have smaller $L_{uu}$. These regions of reduced correlation length coincide with areas of elevated turbulence intensity observed in Figure \ref{fig:channel_stressconts}. Beyond approximately $y/\delta \approx 0.6$, the spanwise variation diminishes and $L_{uu}$ settles at a nearly constant value similar to that of a smooth-wall flow (not shown for brevity). For $Re_\tau \approx 2000$ and $Re_\tau \approx 3000$ the correlation lengths are very similar, whereas at $Re_\tau \approx 4000$ the values are slightly larger --- potentially indicating subtle structural changes at higher Reynolds numbers and this is also observed in the orientation of the elliptical contours (i.e. tilt of the correlation contour - not shown for brevity). Overall, the slight Reynolds number dependence observed here for $S/W = 2$ case is absent in $S/W = 4$. This points to subtle structural changes at higher Reynolds numbers for surfaces with strong secondary flows and should be further explored in future studies.

\section{Conclusions}\label{sec:conclusions}

This study investigated the Reynolds number effect on secondary flows generated by two different ridge-type configurations in channel flow. To our knowledge, these experiments provide the first high-$Re$ channel flow data on ridge-induced secondary flows. The global skin friction measurements indicate that the the Hama roughness function, $\Delta U^+$, computed with skin-friction data, for these surfaces falls within the transitionally rough regime typical of homogeneous roughness up to $Re_\tau \approx 2000$.  At higher Reynolds numbers ($Re_\tau > 3000$), the flow appears to approach the fully rough asymptote (commonly associated with fully rough homogeneous rough surfaces) at low values of $\Delta U^+$ (of approximately 3). This is the first indication that these ridge-like surfaces could reach ``fully-rough'' behaviour in channel flows as most previous work for different types of ridges have only observed transitionally rough behaviour. 

Analysis of the mean flow and turbulent statistics reveals that while the presence of ridge-induced secondary flows modulates the mean velocity, Reynolds stresses and dispersive stresses, their spatial distribution and overall magnitude remain largely invariant with increasing $Re_\tau$. Notably, the $S/W=2$ configuration exhibits stronger stress concentrations and a broader spatial influence compared to the $S/W=4$ case, which likely contributes to its higher drag. Nonetheless, the combined Reynolds and dispersive stress fields in the outer region converge towards the behaviour of smooth-wall channel flow for both surfaces. Two-point correlation analysis further suggests that the structure, spatial extent and magnitude of the secondary flows are only marginally affected by Reynolds number. Near the wall, turbulent structures exhibit smaller correlation lengths in regions of strong secondary flow (both HMP and LMP regions). In contrast, far from the wall the flow becomes nearly homogeneous, with correlation lengths approaching those typical of smooth-wall flows. The orientation of the structures aligns with the radial motions of the secondary flows and increases slightly with Reynolds number. Moreover, the $S/W=2$ configuration exhibits stronger secondary flow effects --- shown by smaller $L_{uu}$ and larger $\theta$ --- than the $S/W=4$ case.

In summary, the experimental results demonstrate that ridge-induced secondary flows significantly modulate near-wall turbulence while leaving the outer flow largely unaffected by changes in Reynolds number. These findings enhance our understanding of how surface heterogeneity influences drag and turbulent structure in channel flows and provide a basis for further investigation into the mechanisms behind ridge-induced flow modifications.

\begin{acknowledgments}
We gratefully acknowledge the financial support from EPSRC (Grant Ref no: EP/V00199X/1).

\end{acknowledgments}

\bibliography{UOS.bib}

\end{document}